\documentclass[
  journal=pasa,
  manuscript=research-paper, 
  year=2020,
  volume=37,
]{cup-journal}
\usepackage{hyperref}
\usepackage{bbm}
\usepackage{microtype,siunitx,booktabs}
\title{Bayesian evidence-driven diagnosis of instrumental systematics for sky-averaged 21-cm cosmology experiments}

\author{K. H. Scheutwinkel}
\affiliation{Astrophysics Group, Cavendish Laboratory, J. J. Thomson Avenue, Cambridge CB3 0HE, UK}
\email[K. H. Scheutwinkel]{khs40@cam.ac.uk}

\author{E. de Lera Acedo}
\affiliation{Astrophysics Group, Cavendish Laboratory, J. J. Thomson Avenue, Cambridge CB3 0HE, UK}
\alsoaffiliation{Kavli Institute for Cosmology, Madingley Road, Cambridge CB3 0HA, UK}

\author{W. Handley}
\affiliation{Astrophysics Group, Cavendish Laboratory, J. J. Thomson Avenue, Cambridge CB3 0HE, UK}
\alsoaffiliation{Kavli Institute for Cosmology, Madingley Road, Cambridge CB3 0HA, UK}

\received {dd Mmm YYYY}
\revised  {dd Mmm YYYY}
\accepted {dd Mmm YYYY}
\published{dd Month YYYY}

\keywords{dark ages, reionisation, first stars -- methods: statistical -- methods: data analysis} 

\begin{document}

\begin{abstract}
We demonstrate the effectiveness of a Bayesian evidence-based analysis for diagnosing and disentangling the sky-averaged 21-cm signal from instrumental systematic effects. As a case study, we consider a simulated REACH pipeline with an injected systematic. We demonstrate that very poor performance or erroneous signal recovery is achieved if the systematic remains unmodelled. These effects include sky-averaged 21-cm posterior estimates resembling a very deep or wide signal. However, when including parameterised models of the systematic, the signal recovery is dramatically improved in performance. Most importantly, a Bayesian evidence-based model comparison is capable of determining whether or not such a systematic model is needed as the true underlying generative model of an experimental dataset is in principle unknown. We, therefore, advocate a pipeline capable of testing a variety of potential systematic errors with the Bayesian evidence acting as the mechanism for detecting their presence.
\end{abstract}

\section{INTRODUCTION }
With 21-cm cosmology \citep{furlanetto_cosmology_2006}, we can potentially probe the earliest phases of the Universe after the cosmic microwave background (CMB) photons decoupled from the dense plasma so that protons and electrons could recombine to form neutral hydrogen when it was energetically favoured. Neutral hydrogen can absorb or emit the hyperfine HI line with a characteristic wavelength of $\lambda = 21$ cm. Finding the 21-cm line in emission or absorption state is dependent on the spin temperature defined through the relative occupancy rates between the excited and neutral state of hydrogen. When the spin temperature is lower than the background radiation in the universe, we will find the 21-cm signal in ``absorption'' and vice versa for emission. This temperature difference is known as the sky-averaged 21-cm signal and it is predicted to have a characteristic absorption feature \citep{pritchard_evolution_2008} which is evolving during Cosmic Dawn (CD) and the Epoch Of Reionisation (EOR). Physical effects affecting the signal shape include the Wouthuysen-Field effect \citep{wouthuysen_excitation_1952, field_excitation_1958} of Lyman-$\alpha$ photon coupling, X-Ray heating through high-energetic X-ray photons, and the progressive reionisation of the hydrogen due to star-forming regions at EOR. For a detailed review of the field see \cite{furlanetto_cosmology_2006, pritchard_21_2012, morales_reionization_2010, liu_data_2020}. 

Observatories trying to detect the 21-cm hydrogen line are designed in one of two ways. First, the interferometric approach such as HERA \citep{deboer_hydrogen_2017}, SKA \citep{dewdney_square_2009}, LOFAR \citep{van_haarlem_lofar_2013}, MWA \citep{tingay_murchison_2013} attempting to characterise the spatial fluctuations of the 21-cm hydrogen line. Second, the simpler approach uses a wide-beam single antenna system, thus averaging over the whole sky to measure the integrated emission of bright and faint radio objects. However, they are harder to calibrate as opposed to multi-antenna systems e.g. the phase shift calibration of a single antenna, and lower resolving power to detect millikelvin-level signals. These measurements are known as the sky-averaged or global signal experiment \citep{liu_global_2013} with observatories such as EDGES \citep{bowman_toward_2008}, SARAS 2 and 3 \citep{singh_saras_2018-1, singh_detection_2022}, LEDA \citep{price_design_2018}, PRIZM \citep{philip_probing_2018}, REACH \citep{de_lera_acedo_reach_2022} and many more in the low radio frequency regime to probe the CD and EOR.

Researchers from EDGES \citep{bowman_absorption_2018} were the first to claim a detection of an absorption profile at CD with a flattened Gaussian shape, centred at $f_{0,21} = 78 \pm 1$ MHz with an amplitude of $A_{21} = 500^{+500}_{-200}$ mK. This result is in tension with the astrophysical models of \citep{cohen_charting_2017}, who postulated a Gaussian-like profile with a maximum amplitude of around $A_{21} \approx 240$ mK. Consequently, candidates to explain these results include new physics, where we either have excessive cooling of the IGM beyond the adiabatic limit involving dark-matter particle interaction \citep{barkana_possible_2018, barkana_signs_2018, munoz_insights_2018} or an enhanced radio background \citep{jana_radio_2019, fialkov_signature_2019, mirocha_what_2019} with strong implications on high-redshift star-formation \citep{mittal_implications_2022}.

However, there is also discussion that does not involve new physics i.e. an astrophysical origin of the signal. Explanations could include instrumental effects such as a ground plane artefact \citep{bradley_ground_2019} mimicking an absorption profile or general concerns about the data analysis. A re-examination of the EDGES result lead by \cite{hills_concerns_2018} (H18) concluded that the best fitting profile reported by EDGES is not a unique solution and contains unphysical values in the ionospheric foreground model with the possibility of an uncalibrated 12.5 MHz sinusoidal structure in the data. This ignited an open scientific debate as \cite{bowman_reply_2018} argued that H18 concerns about unphysical parameters for the ionospheric model stems from covariance between ionospheric and astronomical foreground parameters where small errors in the latter and residual effects during calibration processes can result to biased recovered ionospheric parameters. However, they state that it is possible to extract a sky-averaged 21-cm signal without an absolutely calibrated foreground component as only relatively calibrated frequency channels are needed. Moreover, they acknowledge the ambiguous solution of the recovered shape, however, state that a general absorption profile is a priori most physically justifiable.

Another re-examination of the EDGES low-band dataset lead by \cite{sims_testing_2020} conducted an analysis by including a sinusoidal systematic model and concluded that the models that account for these structures are preferred over models that do not include them. However, they could not clearly distinguish whether there is a clear preference for a sky-averaged 21-cm signal model over models that do not include the signal. Moreover, with other data analysis approaches such as applying a different foreground model using maximally smooth functions (MSF) \citep{singh_redshifted_2019, bevins_maxsmooth_2021}, one can also detect unmodelled sinusoids in the residuals of the EDGES low-band dataset, which are also similarly seen in another sky-averaged experiment LEDA \citep{price_design_2018} and SARAS 2 \citep{bevins_comprehensive_2022}.
Finally, data obtained by the SARAS 3 radiometer reject the presence of the best-fitting absorption feature reported by EDGES with 95.3 \% confidence \citep{singh_detection_2022}.

With this work, we investigate the case of systematic effects i.e. a non astrophysical origin for the Bayesian pipeline of REACH. We introduce unmodelled sinusoidal structures inside forward modelled antenna temperature datasets, and study its influence on the resulting sky-averaged 21-cm signal parameter estimation using a Bayesian inference framework. In Section \ref{sec:BayesianInf}, we introduce the Bayesian inference framework, specifically the parameter estimation and model comparison component.
In Section \ref{sec:ForwardModel}, we describe how we generated our antenna temperature datasets using a physically motivated forward model.
In Section \ref{sec:BayesModel}, we discuss how to statistically model the antenna temperature datasets in a Bayesian way using a likelihood function and assigning a prior distribution on the model parameters. In Section \ref{sec:Results}, we quantify our results by using Bayesian evidence-driven model selection and  a Goodness-of-Fit test and in Section \ref{sec:Conclusion} we conclude and summarise our results.

 \section{BAYESIAN INFERENCE AND NESTED SAMPLING}
\label{sec:BayesianInf}
To analyse the dataset we use Bayesian inference \citep{sivia_data_2006}, a statistical modelling framework where we must assign a prior distribution $p(\theta|M)$ for a model $M$ with its parameters $\theta$ to recover the posterior distribution $p(\theta|D, M)$ of the parameters after the dataset $D$ has been observed. This is achieved by applying Bayes theorem:
\begin{equation}
    p(\theta|D,M) = \frac{p(D|\theta, M) p(\theta|M)}{p(D|M)},
\end{equation}
where $L(\theta) \equiv p(D|\theta, M)$ is the likelihood function of the model parameters and $\mathcal{Z} \equiv p(D|M)$ the Bayesian evidence.

To recover the posterior distribution for parameter estimation and the Bayesian evidence $\mathcal{Z}$ for model comparison, we use nested sampling \citep{skilling_nested_2006}. With nested sampling, one computes the Bayesian evidence by gradually shrinking the prior volume fraction $dX$: \begin{equation}
    \mathcal{Z} = \int \mathcal{L}(\theta) p(\theta|M) d\theta = \int_0^1 \mathcal{L}(X) dX,
\end{equation}
through sampling from the prior which have likelihoods higher than an evolving likelihood constraint $\mathcal{L}$:
\begin{equation}
    X(\mathcal{L}) = \int_{\mathcal{L}(\theta) > \mathcal{L}}  p(\theta|M) d\theta.
\end{equation}
In practice, we use $\texttt{PolyChord}$ \citep{handley_polychord_2015, handley_polychord_2015-1} to find samples subject to the likelihood constraint $\mathcal{L}(\theta) > \mathcal{L}$ which implements slice sampling to draw new proposal samples. As \texttt{PolyChord} is a sampling-based algorithm, one can use the computed Bayesian evidence $\mathcal{Z}$ to get posterior samples $\theta^*$, therefore tackling the parameter estimation and model comparison aspects of Bayesian inference simultaneously. 

To compare competing models we need to apply Bayes Theorem on the Bayesian evidence $\mathcal{Z}$: \begin{equation}
\label{eqn:modelprob}
    p(M|D) = \frac{p(D|M)p(M)}{p(D)},
\end{equation}
to recover the model probabilities $p(M|D)$. We compare competing models $M_1$ and $M_2$ by forming the logarithmic Bayes factor: \begin{equation}
\label{eqn:BayesFactor}
    \log \mathcal{K} = \log p(M_1|D) - \log p(M_2|D),
\end{equation}
where we assume equal prior model probabilities: $p(M_1) = p(M_2)$. A positive Bayes factor $\log \mathcal{K} > 0$ indicates a model preference of $M_1$ over the competing model $M_2$.
\section{FORWARD MODELLING THE DATASET $D$}
\label{sec:ForwardModel}
\begin{figure}
    \centering
    \includegraphics{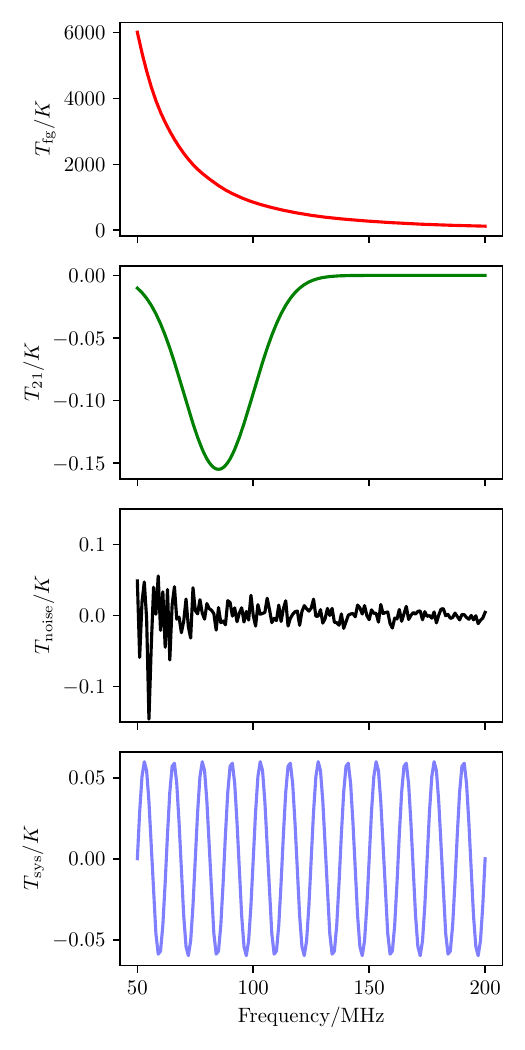}
    \caption{Dataset composition, from top to bottom: the foreground $T_{\mathrm{fg}}$ contribution, the Gaussian sky-averaged 21-cm absorption signal $T_{21}$, the heteroscedastic radiometric noise $T_{\mathrm{noise}}$, a sinusoidal systematic structure $T_{\mathrm{sys}}$.}
    \label{fig:dataset}
\end{figure}

To generate the dataset $D$ for this analysis, the forward model splits the process into a sum of physically motivated components:
\begin{equation}
\label{eqn:dataset}
    D \equiv T_{\mathrm{data}}= T_{\mathrm{fg}} + T_{21}+ T_{\mathrm{sys}}+ T_{\mathrm{noise}},
\end{equation}
with $T_{\mathrm{fg}}$ the simulated foreground, $T_{21}$ the sky-averaged 21-cm signal, $T_{\mathrm{sys}}$ the systematic structure and $T_{\mathrm{noise}}$ the noise component of the antenna temperature. The decomposition can be seen in Figure (\ref{fig:dataset}) and the next sections describe how we simulate each component.

\subsection{Sky simulation}
To simulate the foreground data we use the foreground modelling framework of \cite{anstey_general_2021} that is used as the standard pipeline of REACH \citep{de_lera_acedo_reach_2022}. The pipeline utilise the 2008 Global Sky Model (GSM \cite{de_oliveira-costa_model_2008}) at 408 MHz, $T_{408}(\Omega)$ and 230 MHz, $T_{230}(\Omega)$ and construct the spatially varying spectral index:
\begin{equation}
\label{eqn:specMap}
    \beta(\Omega) = \frac{\log \left(\frac{T_{230}(\Omega) - T_{\mathrm
    {CMB}}}{T_{408}(\Omega) - T_{\mathrm
    {CMB}}}\right)}{\log \left(\frac{230}{408} \right)},
\end{equation}
with $T_{\mathrm{CMB}}$ the cosmic microwave background at $T_{\mathrm{CMB}} = 2.725$ K. This spatially variable spectral index is a physically motivated choice to realistically model the spatial power distribution pattern through its spectral index. Hence, assuming a simple constant spectral index $\beta(\Omega) = \beta_0 = \mathrm{const}$ is an unfavourable and unphysical choice in the simulation process. With this spectral index and $T_{230}(\Omega)$ we simulate the foreground component:
\begin{equation}
    T_{\mathrm{sim}}(\nu, \Omega) = (T_{230} - T_{\mathrm{CMB}} ) \left(\frac{\nu}{230} \right)^{-\beta(\Omega)} + T_{\mathrm{CMB}},
\end{equation}
with $T_{\mathrm{CMB}} = 2.725$ K the cosmic microwave background. We choose $T_{230}(\Omega)$ as the base map as it is an approximated ``empty'' 50-200 MHz representation of the sky without having to worry about contamination introduced by the sky-averaged 21-cm signal as we will add it later on.
We convolve this simulated foreground map with the beam pattern $B_{\Omega}$ of a conical log-spiral antenna \citep{dyson_characteristics_1965}:

\begin{equation}
    T_{\mathrm{fg}}(\nu) =\frac{1}{4 \pi} \int_{\Omega} B(\Omega,\nu) T_{\mathrm{sim}}(\Omega,\nu) d\Omega.
\end{equation}

We choose the conical log-spiral antenna beam pattern as \cite{anstey_informing_2021} reported that this beam pattern has the best performance for correctly extracting a sky-averaged 21-cm signal out of five antenna designs.

The conical log-spiral antenna is FEKO simulated \citep{elsherbeni_antenna_2014}, assumed to be in free-space and located in ideal non-reflective conditions from the ground. We note that this setup is not physically realistic as it does not introduce instrumentally caused systematics, however, due to these ideal conditions we can study how an synthetically added systematic structure is generally handled by the pipeline. One could expand the simulations by adding a ground plane or soil beneath the antenna for subsequent research.
The initialisation of this sky-averaged experimental setup is located at the Karoo Radio Reserve in South Africa,
at -30.71131° latitude and 21.4476236° longitude where REACH is being built. For our analysis, we set the observation time to be a snapshot of the sky at `2019-10-01 00:00:00' UTC (LST: 23.99 h) when the Milky Way centre is not at the zenith.
\subsection{Sky-averaged 21-cm signal}
We add a sky-averaged 21-cm absorption signal with a Gaussian shape to the antenna temperature:
\begin{equation}
\label{eqn:GaussianSignal}
    T_{21}(\nu) = -A_{21} \exp\left(-\frac{1}{2 \sigma_{21}^2}(\nu-f_{0,21})^2\right),
\end{equation}
where $\nu$ is the frequency, $A_{21}$ the amplitude of the sky-averaged 21-cm signal, $f_{0,21}$ the central frequency and $\sigma_{21}$ the standard deviation. For the following analysis we set $A_{21} = 155$ mK, $f_{0,21} = 85$ MHz and $\sigma_{21} = 15$ MHz which is an Gaussian approximation of the standard case of the astrophysical models of \cite{cohen_charting_2017}. 
We emphasise here the challenge of the sky-averaged 21-cm signal extraction, the Gaussian absorption signal is in the millikelvin-level, whereas the foreground contribution reaches several thousands of kelvin in our frequency band.

\subsection{Antenna temperature noise}
After the sky has been simulated, we add the noise component $T_{\mathrm{noise}}$. We choose a physically motivated radiometric noise model \citep{kraus_radio_1986} which can be modelled through a Gaussian distribution centred at $T_{\mathrm{fg}}(\nu)$ with heteroscedastic radiometric noise:
\begin{equation}
\label{eqn:radiometricNoise}
    \sigma_{\mathrm{radio}}(\nu) = \frac{\eta T_{\mathrm{fg}}(\nu) + (1-\eta)T_0 + T_{\mathrm{rec}}}{\sqrt{\tau \Delta\nu}},
\end{equation}
with $T_{\mathrm{fg}}$ the antenna temperature, $\eta$ the antenna radiation efficiency, $T_0$ the ambient temperature, $T_{\mathrm{rec}}$ the antenna receiver temperature, $\tau$ the integration time and $\Delta\nu$ the channel width. For our analysis, we choose $\eta = 0.9$, $T_0 = 293.15$ K, $T_{\mathrm{rec}} = 500$ K, $\tau = 10^5$ s and $\Delta\nu = 0.1$ MHz. This choice of parameters suppresses the noise contribution by a factor of $\sqrt{\tau \Delta \nu} = 10^5$ which is the necessary noise level at the lower frequency end, where the antenna temperature $T_{\mathrm{fg}}$ can reach several thousands of kelvins. We note that in a real experimental setup, the integration time $\tau$ might not be constant across the frequency band due to flagging of RFI contamination.

Furthermore, we ``seed'' the noise so that the noise realisation is identical for each dataset. This engenders consistency in the Bayesian evidence when comparing models and consistency of results across datasets.
\subsection{Systematic structure}
\label{subsec:sinusoidal}
For the systematic component, we choose the generalised damped sinusoid:
\begin{equation}
\label{eqn:dampedsinusoidal}
T_{\mathrm{sys}}(\nu) = \left(\frac{\nu}{f_0}\right)^{\alpha} A_{\mathrm{sys}}\sin \left(2\pi \frac{\nu}{P_{\mathrm{sys}}} + \phi_{\mathrm{sys}} \right),
\end{equation}
where $A_{\mathrm{sys}}$ is the amplitude, $\nu$ the frequency, $P_{\mathrm{sys}}$ the period of the sinusoid, $\phi_{\mathrm{sys}}$ the phase, $\alpha$ the damping coefficient and $f_0$ the reference central frequency of the antenna band. For our experimental setup we have the frequency band $50-200$ MHz, therefore $f_0 = 125$ MHz. These types of structures can arise due to cable reflections within an antenna system, or due to frequency-dependent soil-related losses seen as regular or damped sinusoidal structures.

More concisely, such systematic structures can appear in EDGES, LEDA and in SARAS 2 when using foreground modelling techniques such as polynomial or MSFs e.g. in reanalysis done by \cite{hills_concerns_2018, singh_redshifted_2019, bevins_maxsmooth_2021, bevins_comprehensive_2022}. Possible causes could include instrumental effects or a poor foreground modelling. However, given the different approaches of foreground modelling, the similar design and calibration of the sky-averaged 21-cm experiments and the variation of parameters seen e.g. in amplitude and period of the systematics across instruments, an instrumental origin is more probable. Potential causes e.g. in LEDA could be due to direction-dependent gain of the antenna, a band-pass filter that causes frequency-dependent group-delay or rainfall that moisturises the cables and soil  \citep{price_design_2018} or other calibration errors \citep{sims_testing_2020}. Non-instrumental effects such as ionospheric activity, emission from the soil or RFI could also be a possibility \citep{bevins_comprehensive_2022}. This suggest a further investigation of this structure as a potential instrumental systematic with the exact cause still unknown. As the (damped) sinusoidal systematic structure appeared at several independent sky-averaged 21-cm signal experiments, we continue our analysis by artificially injecting this structure into the dataset.

We choose a parameter cube: \\ $A_{\mathrm{sys}} \in [0, 60, 100, 150, 200, 500, 1000]$ mK, \\$P_{\mathrm{sys}} \in [6, 12.5, 25, 50, 100]$ MHz, $\phi_{\mathrm{sys}} \in [0, \frac{\pi}{2}, \pi, \frac{3}{2}\pi]$. For the damping coefficient we set $\alpha \in [0,-2.5]$ where $\alpha = -2.5 $ is similarly seen in the LEDA \citep{price_design_2018} experiment by an analysis using MSF of \cite{bevins_maxsmooth_2021}. It is worth to note that the damping coefficient is in the order of the spectral index of the galactic foregrounds for this frequency band \citep{mozdzen_improved_2017, spinelli_spectral_2021}, suggesting a possibly multiplicative rather than additive systematic. We inject either one of them, damped or not damped systematic, to the dataset such that we create two parameter cubes - one for each systematic structure - where we take the outer product of the parameter vectors.

\section{BAYESIAN MODELLING OF THE DATASET~$D$}
\label{sec:BayesModel}
\subsection{Likelihood function}

Once we have generated the dataset, we define a likelihood function $\mathcal{L}$ for our model and provide prior ranges of the model parameters $\theta$ which \texttt{PolyChord} needs to compute the Bayesian evidence $\mathcal{Z}$ for our chosen model and generate posterior samples $\theta^*$. As we introduce radiometric noise to generate the dataset, we use a radiometric likelihood function:
\begin{equation}
    \label{eqn:radiometricLikelihood}
    \log \mathcal{L} = \sum_{\nu} -\frac12 \log\left(2 \pi \sigma^2_{\mathrm{radio}}(\nu)\right) - \frac12 \left(\frac{T_{\mathrm{data}}(\nu) - T_{\mathrm{M}}(\nu)}{\sigma_{\mathrm{radio}}(\nu)}\right)^2,
\end{equation}
which is a Gaussian likelihood with heteroscedastic radiometric noise of eq. (\ref{eqn:radiometricNoise}) for the dataset $T_{\mathrm{data}}$ and $T_{\mathrm{M}}$ the modelling component.
We define four models:
\begin{equation}
\label{eqn:signalmodel}
    M_1 \equiv T_{M_{1}} =  T_{\mathrm{fg}} + T_{21},
\end{equation} the $\emph{signal model}$ and: 
\begin{equation}
\label{eqn:nosignalmodel}
    M_2 \equiv T_{M_{2}} =  T_{\mathrm{fg}},
\end{equation}
the $\emph{no signal model}$. For both models we leave the systematic structure unmodelled to study its influence on the parameter inference of the sky-averaged 21-cm signal in Sections \ref{sec:BayesFac} and \ref{sec:GoFtest}.

We also introduce the models $M_3$ and $M_4$ where the systematic structure is modelled in addition to the foreground, sky-averaged 21-cm signal and noise component:
\begin{equation}
\label{eqn:signalSinusmodel}
    M_3  \equiv T_{M_{3}}= T_{\mathrm{fg}} + T_{21} + T_{\mathrm{sys}},
    \end{equation} 
that includes the sky-averaged 21-cm signal model and:
\begin{equation}
\label{eqn:signalSinusmodel}
    M_4  \equiv T_{M_{4}}= T_{\mathrm{fg}} + T_{\mathrm{sys}},
    \end{equation}
that does not contain the sky-averaged 21-cm signal. Hence, these models contain a systematic structure that can arise due to experimental setups.
We study the inference results of these models in Section \ref{sec:IncludeModel}.

To model the foreground we use the Bayesian foreground modelling framework with chromaticity correction of \cite{anstey_general_2021}. This model splits the foreground map into $N_{\mathrm{reg}}$ regions of uniform spectral indices to account for chromatic effects. With this subdivided foreground map which is an approximation of the foreground map used for dataset generation, the antenna temperature can be modelled by convolving it with the conical log-spiral beam pattern $B_{\Omega}$. The log-spiral antenna is assumed to be in free-space with no reflections from the ground i.e. we use the same antenna pattern for inference as the antenna pattern used for simulating the dataset. This log-spiral antenna is similarly used as one of the two antennae of REACH.

For our analysis, we set the foreground region parameter to $N_{\mathrm{reg}} = 14$ as complex structure modelling needs generally more foreground region parameters. One could expand this analysis by studying how the inference changes by varying the number of foreground regions. However, we decide to omit this analysis as it introduces another parameter dimension for an already computationally expensive inference. Moreover, our parameter cube of the sinusoidal structure is extensive enough for the purpose of our study. Additionally, \cite{anstey_general_2021} reported that the inference should in principle not differ much but rather be considered as a tool to search the optimum number of foreground regions with the highest Bayesian evidence. Nevertheless, we note that for a real dataset one should incorporate this extra dimension into the analysis. 

For the sky-averaged 21-cm signal component we include a Gaussian signal model of eq. (\ref{eqn:GaussianSignal}) with $\theta_{21} = (f_{0,21 }, \sigma_{21}, A_{21})$ the parameter vector.

For the radiometric noise component, we include the radiometric noise model of eq. (\ref{eqn:radiometricNoise}) through the radiometric likelihood function and parameterise it through \newline $\theta_{\mathrm{noise}} = (T_{\mathrm{rec}}, \eta, \sigma_{\mathrm{noise}})$, where $\sigma_{noise} = 1/\sqrt{\tau \Delta\nu}$ is the noise level.

For the systematic component, we include a parameterised (damped) sinusoidal model of eq. (\ref{eqn:dampedsinusoidal}) $\theta_\mathrm{sys} = (\alpha_{\mathrm{sys}}, A_{\mathrm{sys}}, P_{\mathrm{sys}}, \phi_{\mathrm{sys}})$. We set the damping coefficient $\alpha_{\mathrm{sys}} = 0$ for inference when the dataset does not contain a damped structure.

To summarise, we have the following parameterization for each model: $\theta_{M_1} = (\theta_{\mathrm{fg}}, \theta_{21}, \theta_{\mathrm{noise}})$,  $\theta_{M_2} = (\theta_{\mathrm{fg}}, \theta_{\mathrm{noise}})$,  $\theta_{M_3} = (\theta_{\mathrm{fg}}, \theta_{21}, \theta_{\mathrm{noise}}, \theta_{\mathrm{sys}})$ and  $\theta_{M_4} = (\theta_{\mathrm{fg}}, \theta_{\mathrm{noise}}, \theta_{\mathrm{sys}})$.
\subsection{Prior ranges of parameters}
The prior ranges of the model parameters are listed in Table \ref{tab:Priors} and we note that the parameter grid of the sinusoidal structure of Section \ref{subsec:sinusoidal} exceeds the prior upper ranges of $\sigma_{21}$ and $A_{21}$. We set these upper limits to signal strengths which are similarly seen in \cite{bowman_absorption_2018}. This allows \texttt{PolyChord} to fit for these EDGES-like sky-averaged 21-cm signal shapes within the datasets. The foreground prior limits $\beta_{1:N_{\mathrm{reg}}}$ are the corresponding minimum and maximum values of the spectral index map computed through eq. (\ref{eqn:specMap}). The sinusoidal systematic parameter $A_{\mathrm{sys.}}$ and $P_{\mathrm{sys.}}$ have variable ranges depending on the dataset e.g. 0-0.5 K and 0-50 MHz for datasets with weaker systematic structures versus 0.5 - 1.5 K and 50-150 MHz ranges for datasets with the largest sinusoidal structures. Nevertheless, the ranges remain identical across different models within a dataset.

We note that the resulting analysis is dependent on the choice of prior and their ranges. The prior choice can be optimised if one understands the underlying system well enough such that some prior types and ranges might be more suitable for a given model and system. As we do not know what a reasonably prior can be for such a system, we choose uniform priors to maximise entropy i.e. minimise prior information with ranges that seems most plausible and physically realistic for our analysis with upper signal parameter ranges that include the absorption feature of \cite{bowman_absorption_2018}.
\begin{table}
    \centering
    \begin{tabular}{lll}
\toprule
{}
Parameter &     Type         &          Range  \\
\midrule
$\beta_{1:N_{\mathrm{reg}}}$ & uniform &  2.458-3.146 \\
$f_{0,21}$              &      uniform &     50-200 MHz\\
$\sigma_{21}$            &      uniform &      10-30 MHz \\
$A_{21}$            &      uniform &     0-0.50 K \\
$T_{\mathrm{rec}}$           &  log uniform &   100-1000 K \\
$\eta$             &      uniform &     0.8-1 \\
$\sigma_{\mathrm{
noise}}$           &  log uniform &  $10^{-8}$-0.1 K  \\

$\alpha_{\mathrm{sys.}}$ &       uniform &  0 - 3 \\
$A_{\mathrm{sys.}}$ &       uniform &  variable \\
$P_{\mathrm{sys.}}$ &       uniform &  variable \\
$\phi_{\mathrm{sys.}}$ &       uniform &  0 - 2 $\pi$ \\
\bottomrule
\end{tabular}

    \caption{Prior choices for the foreground spectral indices $\beta_{1:N_{\mathrm{reg}}}$, sky-averaged 21-cm signal shape $f_{0,21}$, $\sigma_{21}$, $A_{21}$, the radiometric noise model $T_{\mathrm{rec}}$, $\eta$, $\sigma_{\mathrm{
noise}}$ and the systematic parameters $\alpha_{\mathrm{sys.}}$, $A_{\mathrm{sys.}}$,$P_{\mathrm{sys.}}$,$\phi_{\mathrm{sys.}}$.}
    \label{tab:Priors}
\end{table}

\subsection{\texttt{PolyChord} settings}
To compute the Bayesian evidence $\mathcal{Z}$ and to recover the posterior distributions of the model through \texttt{PolyChord} we set the following sampling initializations which are listed in Table \ref{tab:PolyChordSettings}.

\begin{table}
    \centering
    \begin{tabular}{ll}
\toprule
{} 
Parameter &      Settings                    \\
\midrule
\texttt{$n_{\mathrm{live}}$} & $N_{\mathrm{dim}}*25$  \\
\texttt{$n_{\mathrm{prior}}$} & $N_{\mathrm{dim}}*25$  \\
\texttt{$n_{\mathrm{fail}}$} & $N_{\mathrm{dim}}*25$  \\
\texttt{$n_{\mathrm{repeats}}$}         &    $N_{\mathrm{dim}}*5$ \\
$\texttt{precision criterion}$           &      0.001 \\
$\texttt{do clustering}$          &      True  \\
\bottomrule
\end{tabular}
    \caption{\texttt{PolyChord} settings with \texttt{$n_{\mathrm{live}}$} the number of live points, \texttt{$n_{\mathrm{prior}}$}the number of initial prior samples before compression, \texttt{$n_{\mathrm{fail}}$} the number of failed spawns before stopping the algorithm and \texttt{$n_{\mathrm{repeats}}$} the number of slice sampling repeats which are all proportional to the model/parameter dimension $N_{\mathrm{dim}}$. The \texttt{precision criterion} is the nested sampling evidence termination criterion and \texttt{do clustering} if clustering of samples should be activated.}
    \label{tab:PolyChordSettings}
\end{table}

Furthermore, we run \texttt{PolyChord} twice where in the second run we use the posterior sample mean ${\bar \theta^*}$ and sample standard deviation $\sigma^*$ of the first run to construct narrower prior ranges $\bar\theta^* \pm 5 \sigma^*$ (of same shape as in Table \ref{tab:Priors}) for the ``enhanced'' run. This allows us to narrow the initial priors and focus the parameter search around the posterior mode engendering a more tightly constrained posterior distribution of the parameters after the second run. A more detailed description of this method can be found in \cite{anstey_general_2021}.

\section{RESULTS}
\label{sec:Results}
\begin{figure*}
    \centering
    \includegraphics{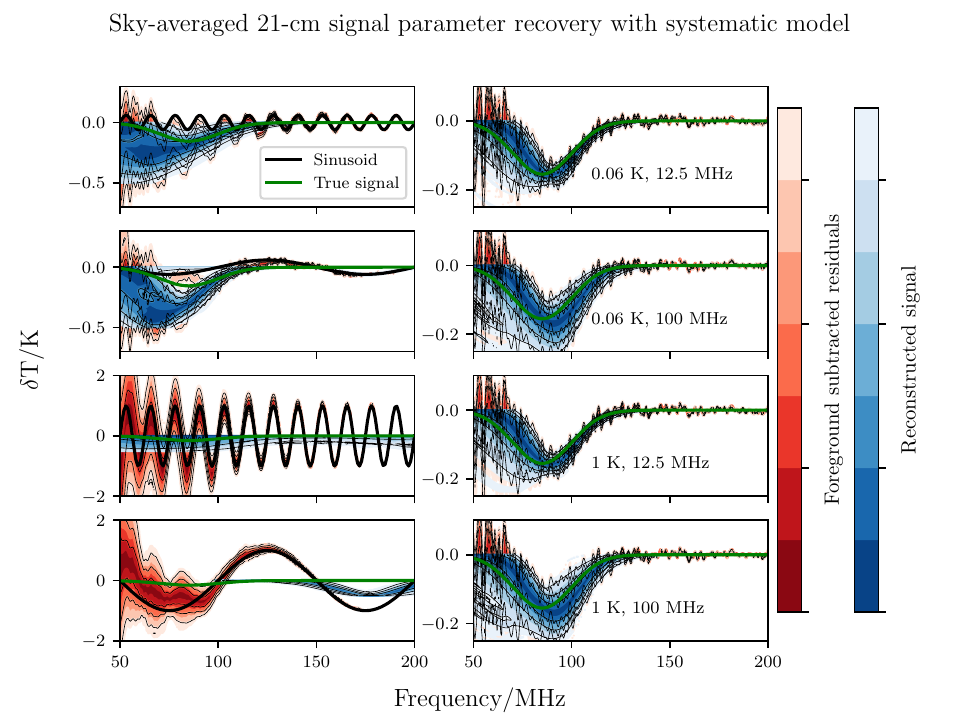}
    \caption{Different sinusoidal structures (black) in the dataset $D$ (parameters in the right panel) and its influence on the foreground subtracted residuals (red), the sky-averaged 21-cm signal inference (blue) in comparison to the true 155 mK Gaussian signal shape (green). The left panel shows the sky-averaged signal recovery when the systematic structure is left unmodelled. The right panel shows the signal recovery when the systematic structure is modelled. One shade in the colorbar represents the $0.5 \sigma$ region.}
    \label{fig:VaryingSinusoids}
\end{figure*}
\begin{figure*}
    \centering
    \includegraphics{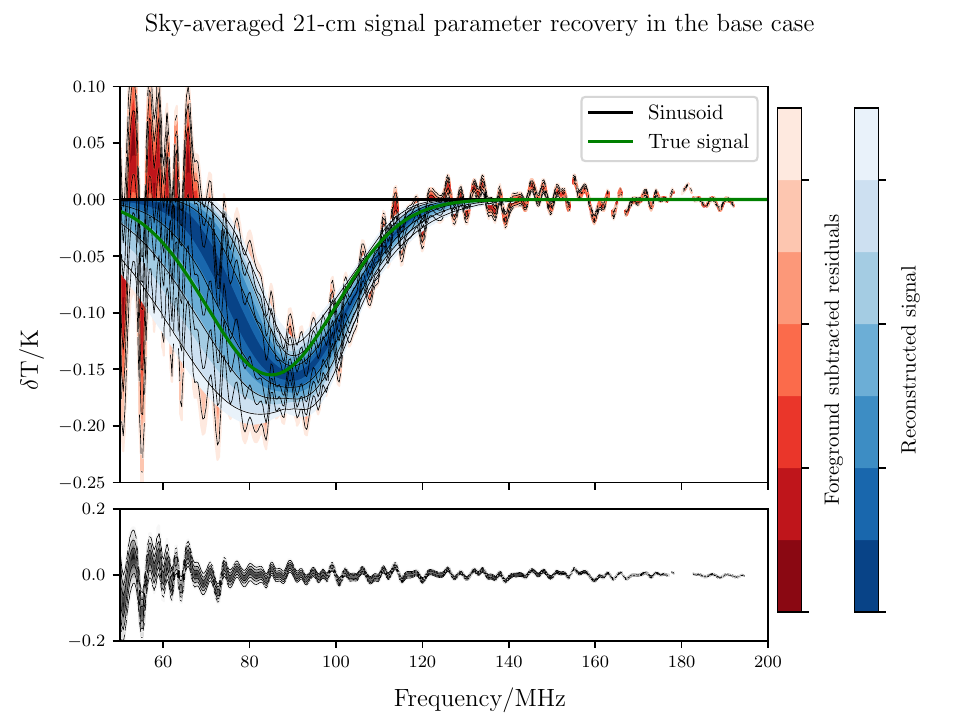}
    \caption{Top: Sky-averaged 21-cm signal recovery (blue) and foreground subtracted residuals (red) when there is no sinusoid present (black line). ``By eye'' the fit looks reasonable compared to the true signal shape (green). 
    Bottom: The radiometric noise residuals. This figure is referred to as the base case.}
    \label{fig:signalRecovery_0mK}
\end{figure*}

In Figures (\ref{fig:VaryingSinusoids}) and (\ref{fig:signalRecovery_0mK}) we present five sky-averaged 21-cm signal extractions when sinusoidal structures without damping are present in the data and are left unmodelled. Each dataset has a varying sinusoidal structure i.e. different parameter pair values $(A_{\mathrm{sys}}, P_{\mathrm{sys}})$ present. Furthermore, these extractions are representative examples of five distinct cases of sinusoidal structures:
\begin{itemize}
    \item low amplitude, low period case, (60 mK, 12.5 MHz),
    \item low amplitude, high period case, (60 mK, 100 MHz),
    \item high amplitude, low period case, (1000 mK, 12.5 MHz),
    \item high amplitude, high period case, (1000 mK, 100 MHz),
    \item base case, no systematic.
\end{itemize}

In all four cases when there is a sinusoid present inside the data, the resulting sky-averaged 21-cm signal shape is highly deformed. The amplitude, scale and central frequency of the sky-averaged 21-cm signal are all affected by the sinusoid i.e. the true Gaussian signal shape is not accurately recovered. We notice a tendency that the extracted sky-averaged 21-cm signal shape has an enhanced amplitude and broadened profile which is similarly reported by EDGES \citep{bowman_absorption_2018} given the predictions of the astrophysical models of \citep{cohen_charting_2017} and the discussion whether or not this is caused by an unmodelled systematic structure seen in the residuals \citep{hills_concerns_2018, bowman_reply_2018, bevins_maxsmooth_2021, singh_redshifted_2019}.

To statistically quantify the signal recovery in a Bayesian way, we will compare these extractions with the inference results of the no-signal model $M_2$ using the Bayesian evidence $\log \mathcal{Z}$ for all the combinations of parameters of the sinusoidal parameter cube in Section \ref{sec:BayesFac}.
We also discuss the Goodness-of-Fit for these five cases of the sinusoid in Section \ref{sec:GoFtest}.
Moreover, when including a systematic model the true sky-averaged 21-cm signal parameters are accurately recovered with the model having the highest Bayesian evidence of all models considered. This is discussed in more detail in Section \ref{sec:IncludeModel}.

\subsection{Bayes Factor $\log \mathcal{K}$ contours}
\label{sec:BayesFac}
\begin{figure*}
    \centering
    \includegraphics{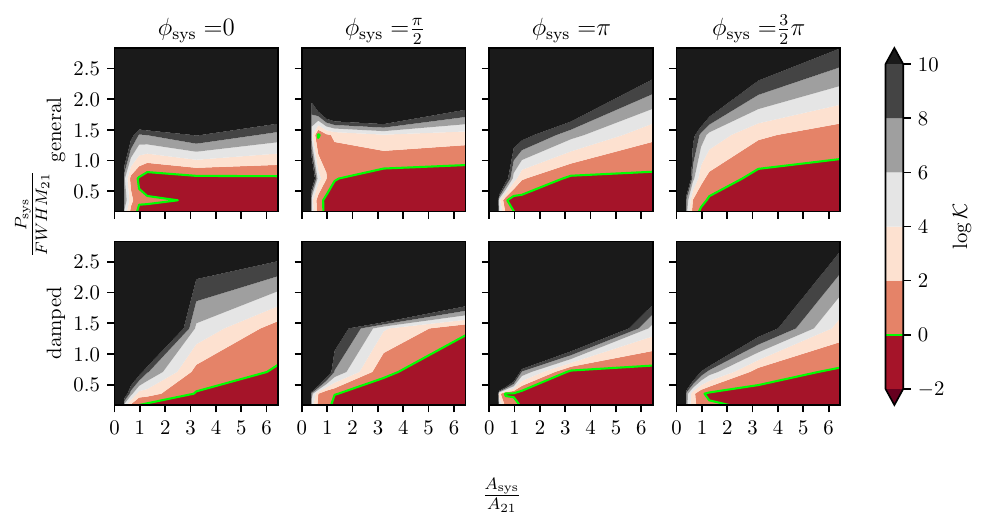}
    \caption{Logarithmic Bayes factor $\log \mathcal{K}$ contour plots for varying parameterisation of the systematic structure and radiometric noise present in the dataset $D$. The damped sinusoid has a damping coefficient $\alpha_{\mathrm{sys}} = -2.5$.}
    \label{fig:logKradioL}
\end{figure*}

For each dataset, we compute the Bayesian evidence $\mathcal{Z}$ of the sky-averaged 21-cm signal model $M_1$ and the no-signal model $M_2$. We then compare these models by forming the Bayes factor $\log \mathcal{K}$ of eq. (\ref{eqn:BayesFactor}). Hence, a positive Bayes factor quantifies that the signal model $M_1$ is preferred for the dataset $D$. 

Figure (\ref{fig:logKradioL}) shows the Bayes factor contour plots for the parameter space $(A_{\mathrm{sys}}, P_{\mathrm{sys}})$ of the sinusoidal systematic relative to the sky-averaged 21-cm signal parameters $A_{21}$ and the Full Width at Half Maximum ($\mathrm{FWHM}_{21}$) for different sinusoidal phases $\phi_{\mathrm{sys}}$.

We find in the high amplitude, low period region of the systematics that the Bayes factor $\log \mathcal{K}$ is preferring the no signal model $M_2$. This is due to the low signal-to-noise ratio between the sky-averaged 21-cm signal and the systematic structure which is also becoming increasingly similar to the radiometric background noise. Hence, Bayesian model selection can not justify the inclusion of a 21-cm signal model and Bayesian model selection penalises more complicated models with constrained parameters that are not necessarily needed, see a discussion in \cite{hergt_bayesian_2021}.

However, this trend seems to break once we have oscillations with higher periods, where the Bayes factor reaches values of $\log \mathcal{K} > 4$. This indicates a strong preference for the 21-cm signal model $M_1$. For these cases the Bayes factor is always preferring the signal model, however, it is not clear if the actual sky-averaged 21-cm signal has been successful. This will be further investigated and discussed in Section \ref{sec:GoFtest}.

In the low amplitude region there is a clear preference for the sky-averaged 21-cm signal model $M_1$, irrespective of the period of the sinusoid. However, as in the high amplitude, high period case, it is not clear whether the sky-averaged 21-cm signal extraction has been successful. 

We note that the model preference stays relatively insensitive to the phase $\phi_{\mathrm{sys}}$ of the systematic. Slight variations e.g. an enhanced preference of the no-signal model $M_2$ for  $\phi_{\mathrm{sys}} = 3/2 \pi$ is due to the varying superposing effects between the amplitude of the sinusoidal and the amplitude of the 21-cm signal.

A similar plot has been generated for the frequency decreasing damped sinusoidal structure which is also shown in Figure (\ref{fig:logKradioL}). Here we observe the same Bayes factor trends as the general sinusoidal structure and it is also not clear whether the recovery of the sky-averaged 21-cm signal was successful.

\subsection{Goodness-of-Fit test}
\label{sec:GoFtest}

\begin{figure*}
    \centering
    \includegraphics{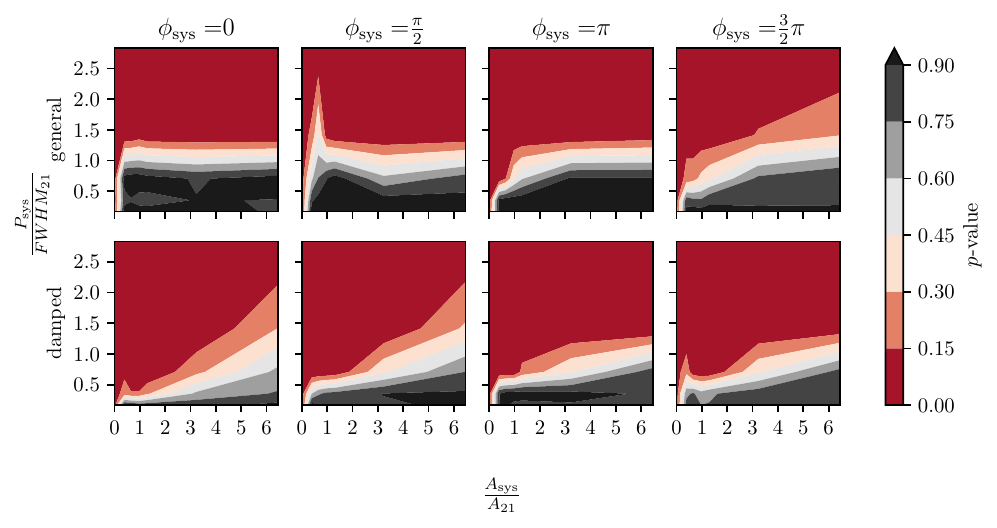}
    \caption{Goodness-of-Fit test $p$-value contour plots with the presence of a systemic structure and radiometric noise in the dataset $D$. The damped sinusoid has a damping coefficient $\alpha_{\mathrm{sys}} = -2.5$.}
    \label{fig:pvalueContours}
\end{figure*}

\begin{figure*}
    \centering
    \includegraphics{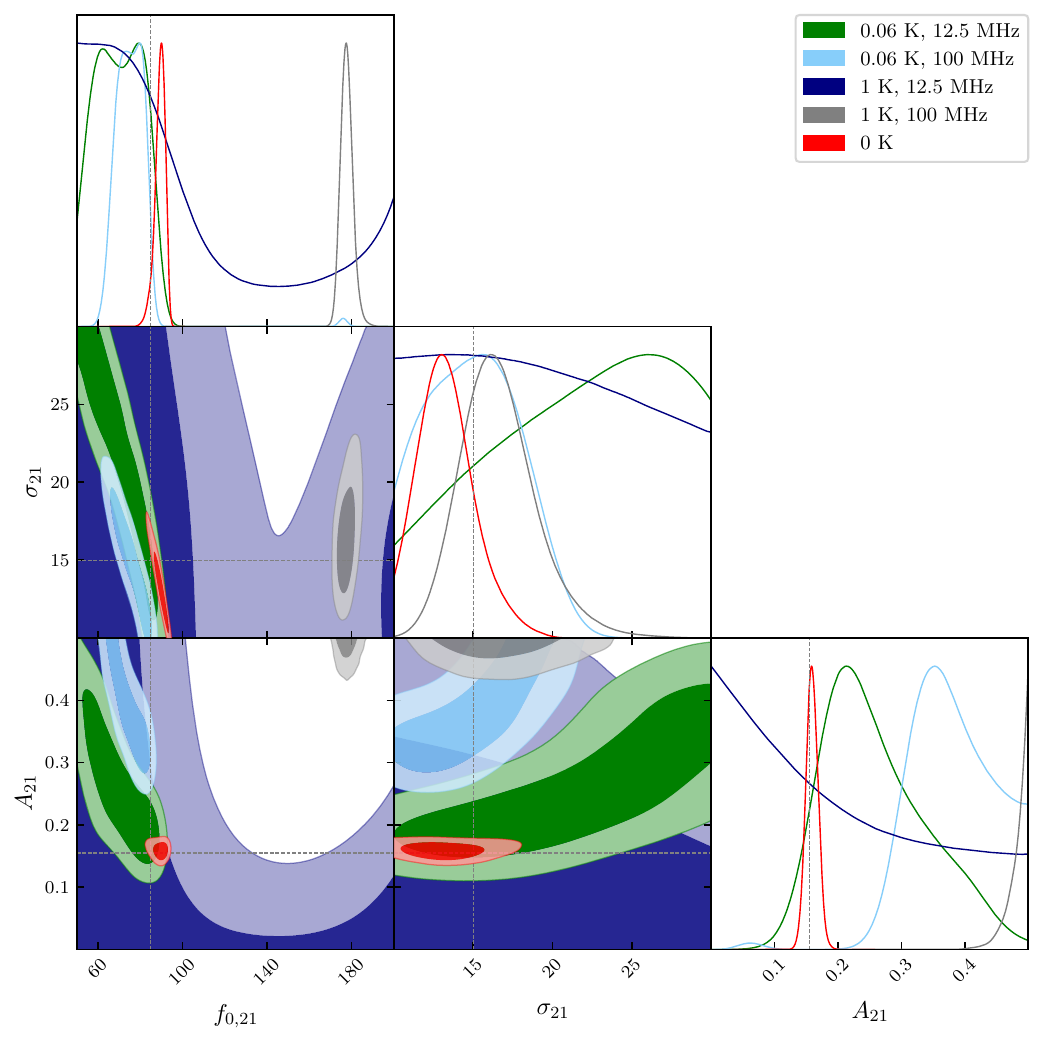}
    \caption{Marginalised posterior distributions for all signal extractions when an unmodelled systematic structure is inside the dataset. The posterior distribution of the base case is in red. The black dotted lines represent the true parameters.}
    \label{fig:marginPost}
\end{figure*}

The Bayes factor analysis is the Bayesian way of quantifying which competing model is preferred, however, this analysis does not quantify if the extraction of the sky-averaged 21-cm signal has been successful i.e. the true sky-averaged 21-cm signal shape has been recovered. 
To quantify the parameter recovery we use a Goodness-of-Fit test by constructing a $p$-value that quantifies how well the posterior distribution is recovering the true signal parameters. Then, we will put that parameter recovery in perspective to the compression rate of the prior as they are related through the Occam's equation of the Bayesian evidence \citep{hergt_bayesian_2021} and therefore contributing to the Bayes factor.
To construct the $p$-value, we need to evaluate the marginalised posterior probabilities of the true signal parameters and the posterior samples. To estimate the probabilities, we learn the marginalised sky-averaged 21-cm signal posterior distribution through normalizing flows \citep{kobyzev_normalizing_2021, papamakarios_normalizing_2021}. 

Normalizing flows are a density estimation method that utilise a series of bijective transformations to find an estimate $\hat P$ of a probability density function. More complex methods such as masked autoregressive flows \citep{papamakarios_masked_2018} utilise a neural network as a bijective transformation and are starting to be applied in cosmology \citep{alsing_nested_2021} to learn sample distributions.

In practice, we use \texttt{margarine} \citep{2022arXiv220512841B} to train these autoregressive flows on the marginalised posterior distribution of the sky-averaged 21-cm signal parameters and conduct a Goodness-of-Fit test by evaluating the probabilities $\hat P(\theta|D)$ of the true signal parameters $\theta_{\mathrm{true}}$ and constructing a $p$-value:
\begin{equation}
    p = \frac{ \sum_{i=1}^{n} w_i \times \mathbbm{1} \{\hat P(\theta^*_i|D) < \hat P(\theta_{\mathrm{true}}|D) \}}{\sum_{i=1}^{n} w_i},
\end{equation}
where $\mathbbm{1}$ is the indicator function for our posterior samples $\theta^*_i$ with weights $w_i$. With this definition, a $p$-value of 0 indicates that no posterior sample has lower probability than the true signal parameters, and p = 1 would indicate that every posterior sample has lower probability than the true signal parameters. This validation technique of posterior distributions can be found in more detail in \cite{harrison_validation_2015}. 

We generated an equivalent contour plot of $p$-values using our parameter grid $(A_{\mathrm{sys}}, P_{\mathrm{sys}})$ with phase variations $\phi_{\mathrm{sys}}$ of the sinusoidal structure in Figure (\ref{fig:pvalueContours}) for the general and frequency decreasing damped cases. We also show the marginalised posterior distributions of the recovered sky-averaged 21-cm signal parameters in Figure (\ref{fig:marginPost}). The $p$-value contour plots can similarly be divided into two categories of sinusoidal systematics, the high or low period case.

For the base case where there is no systematic structure present inside the data ($A_{sys} = 0$ mK), we observe a $p$-value of $p \approx 0.07$. This is due to the true parameters for the sky-averaged 21-cm signal parameter pair ($\sigma_{21}, f_{0,21}$) being outside of the tightly constrained light red shaded $2\sigma$ region of the marginalised posterior distribution shown in Figure (\ref{fig:marginPost}) which is attributed to the approximated foreground model we used in the modelling process. Nevertheless, we will consider this fit and $p$-value as well recovered given the compression rate of the prior to the posterior and the Bayes factor showing a preference for the signal model.

For the high period cases of the sinusoidal structure, we find $p\approx 0$, therefore, the true signal parameters have unlikely been generated by the posterior distribution of the samples. Combining these findings with the Bayes factor contour plots, we can infer that the pipeline is able to extract a sky-averaged 21-cm signal but not the correct one, as these large oscillations are mimicking a sky-averaged 21-cm signal towards the higher frequency end where the foreground and noise contribution is weaker which is seen in the marginalised signal estimates.

For the lower period cases, we find that the $p$-values are significantly higher than the base case when there is no systematic structure in the dataset. In this region, the systematic structure is superposing with the 21-cm signal, i.e. the sinusoidal structure is fitted as part of the 21-cm signal model engendering biased $p$-values. This can be seen in the marginalised posterior distribution, where the true signal parameter are within the contours and not as dominated by the systematic e.g. through major shifting of the central frequency $f_{0,21}$ or deeply enhancing the amplitude $A_{21}$. Additionally, the signal-to-noise ratio is also very low hence showing weaker signs of compression relative to the prior, therefore allowing a wider signal parameter range to fit these datasets. This is consistent with the Bayes factor analysis that shows in this parameter region the no signal model $M_2$ with fewer parameters is preferred. Hence, this region is unsuitable for successful sky-averaged 21-cm signal extraction if the systematics are left unmodelled.

Only when decreasing the amplitude and the period of the sinusoidal significantly $\frac{A_{\mathrm{sys}}}{A_{21}} \leq 0.5$, $\frac{P_{\mathrm{sys}}}{FWHM_{21}} \leq 0.5$, the Bayes factor and the $p$-value indicate a well recovered fit but it is still higher than in the base case. In this regime, the systematic structure is also fitted as part of the 21-cm signal model resulting in a higher $p$-value and biased parameter mean estimates. We conclude that in this parameter region the sky-averaged 21-cm signal recovery can be partly successful but at the cost of a deformed sky-averaged 21-cm signal due to higher parameter variances that do not accurately resemble the true sky-averaged 21-cm signal shape, as seen in Figure (\ref{fig:VaryingSinusoids}) top left plot.

Moreover, we see similar trends regarding the $p$-value for the frequency damped sinusoidal structures. However, there is a tendency that the $p$-values are suggesting a better fit when we increase the amplitude of the systematic structure for a given period. This is due to the amplifying effect of the damped systematic amplitude that superposes with the 21-cm signal, which grows exponentially towards lower frequencies. For these damped sinusoidal structures, the pipeline is fitting a signal towards the lower frequency end where the true signal is located therefore suggesting a better fit.

\subsection{Including a systematic model}
\label{sec:IncludeModel}
The previous analysis has shown that an unmodelled systematic structure can have a dramatic influence on the resulting parameter estimation of the sky-averaged 21-cm signal shape. Hence, we will now study the inference when we include a systematic model i.e. examine the dataset with the systematic models $M_3$ and $M_4$.

The recovery of the sky-averaged 21-cm signal shown in Figure (\ref{fig:VaryingSinusoids}) is now more accurate relative to the true sky-averaged 21-cm signal shape for all four cases of $(A_{\mathrm{sys}}, P_{\mathrm{sys}})$ variations. For the high period cases, the central frequency and shape of the sky-averaged 21-cm signal is now more accurately recovered and for the datasets with lower period sinusoids, we see similar success.

\begin{figure*}
    \centering
    \includegraphics{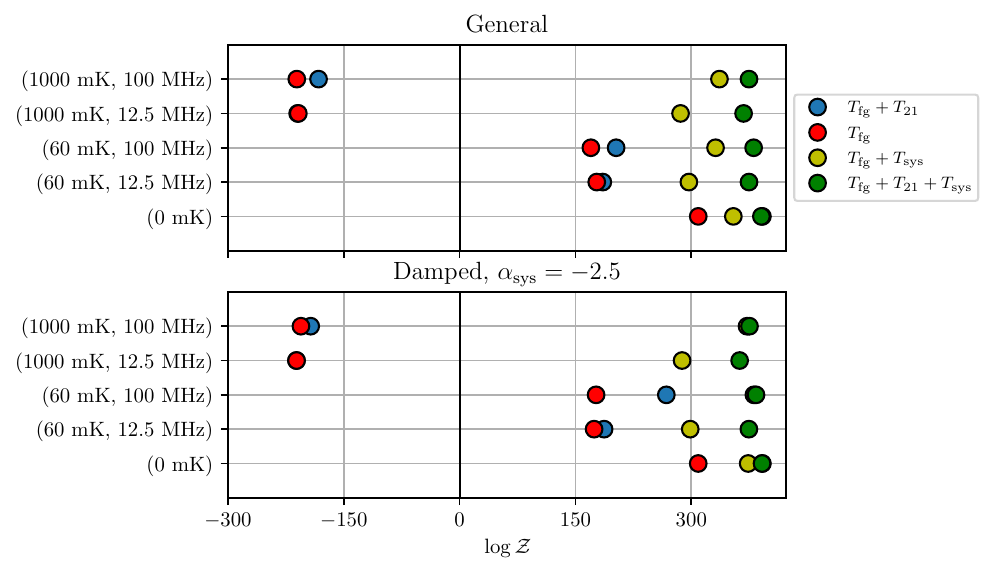}
    \caption{Bayesian evidence $\log \mathcal{Z}$ for four competing models: the signal model (blue), the no signal model (red), the signal model with a sinusoid model (green) and the no signal model with a sinusoid model (yellow). The dataset D includes either general (top) and damped (bottom) sinusoidal structures with varying parameters $(A_{\mathrm{sys}}, P_{\mathrm{sys}})$ for the $\phi_{\mathrm{sys}}= 0$ case. The errors are in the order of $\sigma_{\log \mathcal{Z}} \approx 0.3$.}
    \label{fig:logZComparison}
\end{figure*}

More importantly, the Bayesian evidence (shown in Figure (\ref{fig:logZComparison})) for the true model $T_{\mathrm{fg}} + T_{21} + T_{\mathrm{sys}}$ is the highest out of all four competing models for all systematic datasets considered. Moreover, for the base case when there is no systematic structure inside the dataset, the Bayesian evidence is the highest for the true signal model $T_{\mathrm{fg}} + T_{21}$. Here, the systematic model has a similar but lower Bayesian evidence than the true model. However, the posterior estimate of the amplitude of the systematic is $\bar A^*_{\mathrm{sys}} \approx 0$, therefore, being in agreement with the true model which does not require the systematic model.

Hence, the Bayesian evidence gives us the capability of distinguishing whether we need a systematic model when there is a systematic structure inside the data and correctly preferring the ground truth of this simulated dataset. We note that in a real experimental scenario, we generally do not have access to the true model. However, given our analysis the Bayesian evidence is capable of reliably choosing the models that could be the most likely candidate of the ground truth, therefore providing researcher the tools to continue their research in a given direction by testing a variety of scientifically justifiable systematic models using the Bayesian data analysis pipeline of \cite{anstey_general_2021}.

We note that the for the damped sinusoidal case with high periods, that the Bayes factor is $\log \mathcal{K} \approx 0$ when comparing the models $M_3$ and $M_4$. The corresponding $p$-values of these high period cases are also slightly higher than their low period counterpart. This is due to a larger spread of the posterior samples of the sky-averaged 21-cm signal amplitude $A^*_{21}$. This enhanced uncertainty is caused by the correlation with the phase $\phi_{\mathrm{sys}}$ and period $P_{\mathrm{sys}}$ of the sinusoid. The sinewaves maximum can superpose with the amplitude of the sky-averaged 21-cm signal and due to the long periods affecting the sky-averaged 21-cm signal amplitude more significantly when changing the phase and period marginally. Therefore, the Bayesian evidence can not clearly distinguish whether we need a sky-averaged 21-cm signal model $T_{21}$ due to the larger uncertainty of the sky-averaged 21-cm signal amplitude. 

However, this degeneracy is disentangled for sinusoids with smaller periods where the posterior samples display uncorrelated marginal posterior distributions between sky-averaged 21-cm signal and sinusoidal parameters. Hence, for these low period sinusoids, one achieves successful sky-averaged 21-cm signal recovery and separation of the two model components with the true model having the highest Bayesian evidence.

\section{CONCLUSION}
\label{sec:Conclusion}
Re-examination of the EDGES data analysis has shown that there can be an unmodelled systematic structure present inside the data when using various data analysis techniques such as MSFs or changing the number of polynomial functions. These unmodelled structures are possibly evidence that the reported best-fitting shape of the EDGES collaboration is not purely of astrophysical origin but biased due to instrumental effects of unknown magnitude such as calibration errors. We investigate the influence of these unmodelled systematic structures on the sky-averaged 21-cm signal parameter estimation through the Bayesian inference framework by using a nested sampling-based algorithm \texttt{PolyChord}.

We used a physically motivated forward model to generate datasets $D$ representing the antenna temperature. Each dataset has an identical foreground contribution, radiometric noise and a Gaussian sky-averaged 21-cm absorption signal. After generating these datasets, we added varying sinusoidal structures where we parameterised the amplitude $A_{\mathrm{sys}}$, the period $P_{\mathrm{sys}}$, the phase $\phi_{\mathrm{sys}}$ and the damping coefficient $\alpha_{\mathrm{sys}}$ of the sinusoid. 

We define four models, the signal model $M_1$ with parameters $ \theta_{M_1} = ( \theta_{\mathrm{fg}}, \theta_{21},  \theta_{\mathrm{noise}})$ and the no signal model $M_2$ with parameters  $\theta_{M_2} = ( \theta_{\mathrm{fg}},  \theta_{\mathrm{noise}})$ where we left the systematic structure unmodelled and the systematic models $M_3$ and $M_4$ with the parameterisation $\theta_{M_3} = (\theta_{\mathrm{fg}},  \theta_{21},\theta_{\mathrm{noise}}, \theta_{\mathrm{sys}}),\theta_{M_4} = (\theta_{\mathrm{fg}},\theta_{\mathrm{noise}}, \theta_{\mathrm{sys}}) $, where we include a systematic model. As we used the radiometric noise model to generate our datasets, we use a radiometric likelihood function, which can be modelled through a Gaussian distribution with heteroscedastic radiometric noise. With this radiometric likelihood function and combined with the prior ranges of our parameters, we can recover the marginalised posterior distributions of the parameters using the nested sampling-based algorithm \texttt{PolyChord}. For each dataset, we use \texttt{PolyChord} to generate posterior samples and compute the Bayesian evidence $\log \mathcal{Z}$ for the models considered.

To compare these models, we constructed the logarithmic Bayes factor $\log \mathcal{K}$ by applying Bayes Theorem on the Bayesian evidence $\mathcal{Z}$ to acquire the model probabilities. For our parameterised sinusoids, we computed Bayes factor $\log \mathcal{K}$ contour plots and found that when the systematic structures are left unmodelled the signal model $M_1$ is generally preferred over the no signal model $M_2$ except in the high amplitude, low period regime of the sinusoid. This is due to the low signal-to-noise ratio between the sky-averaged 21-cm signal and the noisy background contributions by the sinusoid and the radiometric noise. 

However, the Bayes factor $\log \mathcal{K}$ contour plots do not contain any information whether the actual sky-averaged 21-cm signal recovery has been successful i.e. the true values of the sky-averaged 21-cm signal are within the posterior sample estimates. To quantify this statistically we used the Goodness-of-Fit test by learning the posterior distribution of our sky-averaged 21-cm signal parameters $\theta_{21} = (f_{0,21}, \sigma_{21}, A_{21})$ through normalizing flows and computed the corresponding $p$-value. We created a $p$-value contour plot for our sinusoidal parameter cube $(A_{\mathrm{sys}}, P_{\mathrm{sys}}, \phi_{\mathrm{sys}})$ similar to the Bayes factor $\log \mathcal{K}$ contours.

This analysis has shown that the sky-averaged 21-cm signal recovery is only slightly successful with high uncertainties in the sky-averaged 21-cm signal parameter estimates for sinusoidal parameter regions where $\frac{A_{\mathrm{sys}}}{A_{21}} < 1$ and $\frac{P_{\mathrm{sys}}}{\mathrm{FWHM}_{21}} < 1$. For the other regions, the corresponding $p$-value was either close to zero or one. A $p$-value of zero is caused by the long period systematic structures mimicking an absorption profile with a tendency towards the higher frequency end of the band. A $p$-value close to one is due to the short period sinusoidal structures being fitted as part of the signal model, engendering biased higher $p$-values. Combining these results with the marginalised posterior distributions and the Bayes factor contour plots, we concluded that these regions are unsuitable for signal recovery.

However, this picture changes dramatically once we add a systematic model next to the foreground, the sky-averaged 21-cm signal and the noise, hence using the models $M_3$ and $M_4$ for inference. We show that for all four regions of $(A_{\mathrm{sys}}, P_{\mathrm{sys}})$ combinations, the resulting sky-averaged 21-cm signal extraction has been successful and the Bayesian evidence $\log \mathcal{Z}$ for the true model $M_3$ is the highest out of all four competing models. Therefore, the Bayesian evidence provides the tools to decide whether we need a systematic model when there is a systematic structure inside the dataset and it is capable of guiding us towards finding the `true' underlying model of a dataset in a real-world scenario.

Furthermore, the marginalised posterior distributions of the parameters show an uncorrelated behaviour between the signal parameters $(f_{0,21}, \sigma_{21}, A_{21})$ and $(A_{\mathrm{sys}}, P_{\mathrm{sys}}, \phi_{\mathrm{sys}})$ for sinusoids with small periods. This indicates a successful separation of these structures in the modelling process, therefore resulting in more precise posterior inferences. For longer periods, there is a tendency that the period and phase of the sinusoid are correlated with the amplitude of the sky-averaged 21-cm signal, hence, making it a crucial task to precisely constrain these sinusoidal parameters through the prior. This is evidence that by including a physically motivated systematic model there is a possibility to do accurate science when there is a systematic structure present in the data.

Overall, this analysis has shown that if there is a systematic structure present inside the dataset, Bayesian inference strongly prefers a model where we include a systematic feature. Additionally, the parameter inference results in a more accurate recovery of the sky-averaged 21-cm signal parameters and more tightly constrained posterior distributions. Finally, this analysis is used to guide the design and the challenging task of calibration of the sky-averaged experiment REACH \citep{de_lera_acedo_reach_2022} where systematic effects can be parameterised, introduced and tested through the Bayesian data analysis pipeline of \citep{anstey_general_2021}. Possible design choices could include the constraint of cable lengths connected to the antenna system that could introduce a varying degree of systematics effects into the data. Finally, this kind of systematics analysis using the Bayesian data analysis pipeline is also applicable to beyond REACH.

\section*{Acknowledgements}
This work was performed using resources provided by the Cambridge Service for Data Driven Discovery (CSD3) operated by the University of Cambridge Research Computing Service, provided by Dell EMC and Intel using Tier-2 funding from the Engineering and Physical Sciences Research Council (capital grant EP/T022159/1), and DiRAC funding from the Science and Technology Facilities Council. KHS would like to thank Dominic Anstey for developing and providing the core Bayesian data analysis pipeline used in REACH and the support of the Hans Werthén foundation of the Royal Swedish Academy of Engineering Sciences. EdLA is supported by the Science and Technologies Facilities Council Ernest Rutherford Fellowship. WH is supported by a Gonville \& Caius College Research Fellowship and Royal Society University Research Fellowship.
\section*{Data availability}
The underlying data is available at Zenodo on \\ \href{https://doi.org/10.5281/zenodo.6474036}{https://doi.org/10.5281/zenodo.6474036}


\bibliography{references}

\begin{thebibliography}{}
\expandafter\ifx\csname natexlab\endcsname\relax\def\natexlab#1{#1}\fi

\bibitem[{Alsing \& Handley(2021)}]{alsing_nested_2021}
Alsing, J., \& Handley, W. 2021, Monthly Notices of the Royal Astronomical
  Society: Letters, 505, L95

\bibitem[{Anstey {et~al.}(2021{\natexlab{a}})Anstey, Cumner, de~Lera~Acedo, \&
  Handley}]{anstey_informing_2021}
Anstey, D., Cumner, J., de~Lera~Acedo, E., \& Handley, W. 2021{\natexlab{a}},
  Monthly Notices of the Royal Astronomical Society, 509, 4679

\bibitem[{Anstey {et~al.}(2021{\natexlab{b}})Anstey, de Lera Acedo, \&
  Handley}]{anstey_general_2021}
Anstey, D., de Lera Acedo, E., \& Handley, W. 2021{\natexlab{b}}, Monthly
  Notices of the Royal Astronomical Society, 506, 2041

\bibitem[{Barkana(2018)}]{barkana_possible_2018}
Barkana, R. 2018, Nature, 555, 71, arXiv: 1803.06698

\bibitem[{Barkana {et~al.}(2018)Barkana, Outmezguine, Redigolo, \&
  Volansky}]{barkana_signs_2018}
Barkana, R., Outmezguine, N.~J., Redigolo, D., \& Volansky, T. 2018, Physical
  Review D, 98, 103005, arXiv: 1803.03091

\bibitem[{Bevins {et~al.}(2022)Bevins, de~Lera~Acedo, Fialkov, Handley, Singh,
  Subrahmanyan, \& Barkana}]{bevins_comprehensive_2022}
Bevins, H. T.~J., de~Lera~Acedo, E., Fialkov, A., {et~al.} 2022, Monthly
  Notices of the Royal Astronomical Society, 513, 4507

\bibitem[{Bevins {et~al.}(2021)Bevins, Handley, Fialkov, de Lera Acedo,
  Greenhill, \& Price}]{bevins_maxsmooth_2021}
Bevins, H. T.~J., Handley, W.~J., Fialkov, A., {et~al.} 2021, Monthly Notices
  of the Royal Astronomical Society, 502, 4405

\bibitem[{{Bevins} {et~al.}(2022){Bevins}, {Handley}, {Lemos}, {Sims}, {de Lera
  Acedo}, {Fialkov}, \& {Alsing}}]{2022arXiv220512841B}
{Bevins}, H. T.~J., {Handley}, W.~J., {Lemos}, P., {et~al.} 2022, arXiv
  e-prints, arXiv:2205.12841

\bibitem[{Bowman {et~al.}(2008)Bowman, Rogers, \& Hewitt}]{bowman_toward_2008}
Bowman, J.~D., Rogers, A. E.~E., \& Hewitt, J.~N. 2008, The Astrophysical
  Journal, 676, 1, arXiv: 0710.2541

\bibitem[{Bowman {et~al.}(2018{\natexlab{a}})Bowman, Rogers, Monsalve, Mozdzen,
  \& Mahesh}]{bowman_absorption_2018}
Bowman, J.~D., Rogers, A. E.~E., Monsalve, R.~A., Mozdzen, T.~J., \& Mahesh, N.
  2018{\natexlab{a}}, Nature, 555, 67, number: 7694 Publisher: Nature
  Publishing Group

\bibitem[{Bowman {et~al.}(2018{\natexlab{b}})Bowman, Rogers, Monsalve, Mozdzen,
  \& Mahesh}]{bowman_reply_2018}
---. 2018{\natexlab{b}}, Nature, 564, E35, number: 7736 Publisher: Nature
  Publishing Group

\bibitem[{Bradley {et~al.}(2019)Bradley, Tauscher, Rapetti, \&
  Burns}]{bradley_ground_2019}
Bradley, R.~F., Tauscher, K., Rapetti, D., \& Burns, J.~O. 2019, The
  Astrophysical Journal, 874, 153, arXiv: 1810.09015

\bibitem[{Cohen {et~al.}(2017)Cohen, Fialkov, Barkana, \&
  Lotem}]{cohen_charting_2017}
Cohen, A., Fialkov, A., Barkana, R., \& Lotem, M. 2017, Monthly Notices of the
  Royal Astronomical Society, 472, 1915, arXiv: 1609.02312

\bibitem[{de~Lera~Acedo {et~al.}(2022)de~Lera~Acedo, de~Villiers, Razavi-Ghods,
  Handley, Fialkov, Magro, Anstey, Bevins, Chiello, Cumner, Josaitis, Roque,
  Sims, Scheutwinkel, Alexander, Bernardi, Carey, Cavillot, Croukamp, Ely,
  Gessey-Jones, Gueuning, Hills, Kulkarni, Maiolino, Meerburg, Mittal,
  Pritchard, Puchwein, Saxena, Shen, Smirnov, Spinelli, \&
  Zarb-Adami}]{de_lera_acedo_reach_2022}
de~Lera~Acedo, E., de~Villiers, D. I.~L., Razavi-Ghods, N., {et~al.} 2022,
  Nature Astronomy, 1, publisher: Nature Publishing Group

\bibitem[{De~Oliveira-Costa {et~al.}(2008)De~Oliveira-Costa, Tegmark, Gaensler,
  Jonas, Landecker, \& Reich}]{de_oliveira-costa_model_2008}
De~Oliveira-Costa, A., Tegmark, M., Gaensler, B.~M., {et~al.} 2008, Monthly
  Notices of the Royal Astronomical Society, 388, 247

\bibitem[{DeBoer {et~al.}(2017)DeBoer, Parsons, Aguirre, Alexander, Ali,
  Beardsley, Bernardi, Bowman, Bradley, Carilli, Cheng, Acedo, Dillon,
  Ewall-Wice, Fadana, Fagnoni, Fritz, Furlanetto, Glendenning, Greig,
  Grobbelaar, Hazelton, Hewitt, Hickish, Jacobs, Julius, Kariseb, Kohn,
  Lekalake, Liu, Loots, MacMahon, Malan, Malgas, Maree, Mathison, Matsetela,
  Mesinger, Morales, Neben, Patra, Pieterse, Pober, Razavi-Ghods, Ringuette,
  Robnett, Rosie, Sell, Smith, Syce, Tegmark, Thyagarajan, Williams, \&
  Zheng}]{deboer_hydrogen_2017}
DeBoer, D.~R., Parsons, A.~R., Aguirre, J.~E., {et~al.} 2017, Publications of
  the Astronomical Society of the Pacific, 129, 045001, arXiv: 1606.07473

\bibitem[{Dewdney {et~al.}(2009)Dewdney, Hall, Schilizzi, \&
  Lazio}]{dewdney_square_2009}
Dewdney, P., Hall, P., Schilizzi, R., \& Lazio, T. 2009, Proceedings of the
  IEEE, 97, 1482

\bibitem[{Dyson(1965)}]{dyson_characteristics_1965}
Dyson, J. 1965, IEEE Transactions on Antennas and Propagation, 13, 488,
  conference Name: IEEE Transactions on Antennas and Propagation

\bibitem[{Elsherbeni(2014)}]{elsherbeni_antenna_2014}
Elsherbeni, A. Z.~a. 2014, Antenna analysis and design using {FEKO}
  electromagnetic simulation software (Edison, NJ : SciTech Publishing, an
  imprint of the IET, [2014] ©2014)

\bibitem[{Fialkov \& Barkana(2019)}]{fialkov_signature_2019}
Fialkov, A., \& Barkana, R. 2019, Monthly Notices of the Royal Astronomical
  Society, 486, 1763, arXiv: 1902.02438

\bibitem[{Field(1958)}]{field_excitation_1958}
Field, G.~B. 1958, Proceedings of the IRE, 46, 240, conference Name:
  Proceedings of the IRE

\bibitem[{Furlanetto {et~al.}(2006)Furlanetto, Oh, \&
  Briggs}]{furlanetto_cosmology_2006}
Furlanetto, S., Oh, S.~P., \& Briggs, F. 2006, Physics Reports, 433, 181,
  arXiv: astro-ph/0608032

\bibitem[{Handley {et~al.}(2015{\natexlab{a}})Handley, Hobson, \&
  Lasenby}]{handley_polychord_2015}
Handley, W.~J., Hobson, M.~P., \& Lasenby, A.~N. 2015{\natexlab{a}}, Monthly
  Notices of the Royal Astronomical Society: Letters, 450, L61, arXiv:
  1502.01856

\bibitem[{Handley {et~al.}(2015{\natexlab{b}})Handley, Hobson, \&
  Lasenby}]{handley_polychord_2015-1}
---. 2015{\natexlab{b}}, Monthly Notices of the Royal Astronomical Society,
  453, 4385, arXiv: 1506.00171

\bibitem[{Harrison {et~al.}(2015)Harrison, Sutton, Carvalho, \&
  Hobson}]{harrison_validation_2015}
Harrison, D., Sutton, D., Carvalho, P., \& Hobson, M. 2015, Monthly Notices of
  the Royal Astronomical Society, 451, 2610

\bibitem[{Hergt {et~al.}(2021)Hergt, Handley, Hobson, \&
  Lasenby}]{hergt_bayesian_2021}
Hergt, L.~T., Handley, W.~J., Hobson, M.~P., \& Lasenby, A.~N. 2021, Physical
  Review D, 103, 123511, arXiv: 2102.11511

\bibitem[{Hills {et~al.}(2018)Hills, Kulkarni, Meerburg, \&
  Puchwein}]{hills_concerns_2018}
Hills, R., Kulkarni, G., Meerburg, P.~D., \& Puchwein, E. 2018, Nature, 564,
  E32, arXiv: 1805.01421

\bibitem[{Jana {et~al.}(2019)Jana, Nath, \& Biermann}]{jana_radio_2019}
Jana, R., Nath, B.~B., \& Biermann, P.~L. 2019, Monthly Notices of the Royal
  Astronomical Society, 483, 5329, arXiv: 1812.07404

\bibitem[{Kobyzev {et~al.}(2021)Kobyzev, Prince, \&
  Brubaker}]{kobyzev_normalizing_2021}
Kobyzev, I., Prince, S.~J., \& Brubaker, M.~A. 2021, IEEE Transactions on
  Pattern Analysis and Machine Intelligence, 43, 3964, conference Name: IEEE
  Transactions on Pattern Analysis and Machine Intelligence

\bibitem[{Kraus {et~al.}(1986)Kraus, Tiuri, Räisänen, \&
  Carr}]{kraus_radio_1986}
Kraus, J.~D., Tiuri, M., Räisänen, A.~V., \& Carr, T.~D. 1986, Radio
  {Astronomy} (Cygnus-Quasar Books), google-Books-ID: KtVFAQAAIAAJ

\bibitem[{Liu {et~al.}(2013)Liu, Pritchard, Tegmark, \& Loeb}]{liu_global_2013}
Liu, A., Pritchard, J.~R., Tegmark, M., \& Loeb, A. 2013, Physical Review D,
  87, 043002, arXiv: 1211.3743

\bibitem[{Liu \& Shaw(2020)}]{liu_data_2020}
Liu, A., \& Shaw, J.~R. 2020, Publications of the Astronomical Society of the
  Pacific, 132, 062001, arXiv: 1907.08211

\bibitem[{Mirocha \& Furlanetto(2019)}]{mirocha_what_2019}
Mirocha, J., \& Furlanetto, S.~R. 2019, Monthly Notices of the Royal
  Astronomical Society, 483, 1980, arXiv: 1803.03272

\bibitem[{Mittal \& Kulkarni(2022)}]{mittal_implications_2022}
Mittal, S., \& Kulkarni, G. 2022, arXiv:2203.07733 [astro-ph], arXiv:
  2203.07733

\bibitem[{Morales \& Wyithe(2010)}]{morales_reionization_2010}
Morales, M.~F., \& Wyithe, J. S.~B. 2010, Annual Review of Astronomy and
  Astrophysics, 48, 127, arXiv: 0910.3010

\bibitem[{Mozdzen {et~al.}(2017)Mozdzen, Bowman, Monsalve, \&
  Rogers}]{mozdzen_improved_2017}
Mozdzen, T.~J., Bowman, J.~D., Monsalve, R.~A., \& Rogers, A. E.~E. 2017,
  Monthly Notices of the Royal Astronomical Society, 464, 4995, aDS Bibcode:
  2017MNRAS.464.4995M

\bibitem[{Muñoz \& Loeb(2018)}]{munoz_insights_2018}
Muñoz, J.~B., \& Loeb, A. 2018, Nature, 557, 684, arXiv: 1802.10094

\bibitem[{Papamakarios {et~al.}(2021)Papamakarios, Nalisnick, Rezende, Mohamed,
  \& Lakshminarayanan}]{papamakarios_normalizing_2021}
Papamakarios, G., Nalisnick, E., Rezende, D.~J., Mohamed, S., \&
  Lakshminarayanan, B. 2021, Journal of Machine Learning Research, 22, 1

\bibitem[{Papamakarios {et~al.}(2018)Papamakarios, Pavlakou, \&
  Murray}]{papamakarios_masked_2018}
Papamakarios, G., Pavlakou, T., \& Murray, I. 2018, arXiv:1705.07057 [cs,
  stat], arXiv: 1705.07057

\bibitem[{Philip {et~al.}(2018)Philip, Abdurashidova, Chiang, Ghazi, Gumba,
  Heilgendorff, Jáuregui-García, Malepe, Nunhokee, Peterson, Sievers, Simes,
  \& Spann}]{philip_probing_2018}
Philip, L., Abdurashidova, Z., Chiang, H.~C., {et~al.} 2018, Journal of
  Astronomical Instrumentation, 08, 1950004, publisher: World Scientific
  Publishing Co.

\bibitem[{Price {et~al.}(2018)Price, Greenhill, Fialkov, Bernardi, Garsden,
  Barsdell, Kocz, Anderson, Bourke, Craig, Dexter, Dowell, Eastwood, Eftekhari,
  Ellingson, Hallinan, Hartman, Kimberk, Lazio, Leiker, MacMahon, Monroe,
  Schinzel, Taylor, Tong, Werthimer, \& Woody}]{price_design_2018}
Price, D.~C., Greenhill, L.~J., Fialkov, A., {et~al.} 2018, Monthly Notices of
  the Royal Astronomical Society, 478, 4193

\bibitem[{Pritchard \& Loeb(2008)}]{pritchard_evolution_2008}
Pritchard, J.~R., \& Loeb, A. 2008, Physical Review D, 78, 103511, arXiv:
  0802.2102

\bibitem[{Pritchard \& Loeb(2012)}]{pritchard_21_2012}
---. 2012, Reports on Progress in Physics, 75, 086901

\bibitem[{Sims \& Pober(2020)}]{sims_testing_2020}
Sims, P.~H., \& Pober, J.~C. 2020, Monthly Notices of the Royal Astronomical
  Society, 492, 22, arXiv: 1910.03165

\bibitem[{Singh \& Subrahmanyan(2019)}]{singh_redshifted_2019}
Singh, S., \& Subrahmanyan, R. 2019, The Astrophysical Journal, 880, 26, arXiv:
  1903.04540

\bibitem[{Singh {et~al.}(2018)Singh, Subrahmanyan, Shankar, Rao, Girish,
  Raghunathan, Somashekar, \& Srivani}]{singh_saras_2018-1}
Singh, S., Subrahmanyan, R., Shankar, N.~U., {et~al.} 2018, Experimental
  Astronomy, 45, 269

\bibitem[{Singh {et~al.}(2022)Singh, Nambissan~T., Subrahmanyan, Udaya~Shankar,
  Girish, Raghunathan, Somashekar, Srivani, \&
  Sathyanarayana~Rao}]{singh_detection_2022}
Singh, S., Nambissan~T., J., Subrahmanyan, R., {et~al.} 2022, Nature Astronomy,
  6, 607, number: 5 Publisher: Nature Publishing Group

\bibitem[{Sivia \& Skilling(2006)}]{sivia_data_2006}
Sivia, D.~S., \& Skilling, J. 2006, Data analysis: a {Bayesian} tutorial, 2nd
  edn., Oxford science publications (Oxford, England: Oxford University Press)

\bibitem[{Skilling(2006)}]{skilling_nested_2006}
Skilling, J. 2006, Bayesian Analysis, 1, 833, publisher: International Society
  for Bayesian Analysis

\bibitem[{Spinelli {et~al.}(2021)Spinelli, Bernardi, Garsden, Greehill,
  Fialkov, Dowell, \& Price}]{spinelli_spectral_2021}
Spinelli, M., Bernardi, G., Garsden, H., {et~al.} 2021, Monthly Notices of the
  Royal Astronomical Society, 505, 1575, arXiv:2011.03994 [astro-ph]

\bibitem[{Tingay {et~al.}(2013)Tingay, Goeke, Bowman, Emrich, Ord, Mitchell,
  Morales, Booler, Crosse, Wayth, Lonsdale, Tremblay, Pallot, Colegate,
  Wicenec, Kudryavtseva, Arcus, Barnes, Bernardi, Briggs, Burns, Bunton,
  Cappallo, Corey, Deshpande, Desouza, Gaensler, Greenhill, Hall, Hazelton,
  Herne, Hewitt, Johnston-Hollitt, Kaplan, Kasper, Kincaid, Koenig,
  Kratzenberg, Lynch, Mckinley, Mcwhirter, Morgan, Oberoi, Pathikulangara,
  Prabu, Remillard, Rogers, Roshi, Salah, Sault, Udaya-Shankar, Schlagenhaufer,
  Srivani, Stevens, Subrahmanyan, Waterson, Webster, Whitney, Williams,
  Williams, \& Wyithe}]{tingay_murchison_2013}
Tingay, S.~J., Goeke, R., Bowman, J.~D., {et~al.} 2013, Publications of the
  Astronomical Society of Australia, 30, e007

\bibitem[{van Haarlem {et~al.}(2013)van Haarlem, Wise, Gunst, Heald, McKean,
  Hessels, de~Bruyn, Nijboer, Swinbank, Fallows, Brentjens, Nelles, Beck,
  Falcke, Fender, Hörandel, Koopmans, Mann, Miley, Röttgering, Stappers,
  Wijers, Zaroubi, van~den Akker, Alexov, Anderson, Anderson, van Ardenne,
  Arts, Asgekar, Avruch, Batejat, Bähren, Bell, Bell, van Bemmel, Bennema,
  Bentum, Bernardi, Best, Bîrzan, Bonafede, Boonstra, Braun, Bregman,
  Breitling, van~de Brink, Broderick, Broekema, Brouw, Brüggen, Butcher, van
  Cappellen, Ciardi, Coenen, Conway, Coolen, Corstanje, Damstra, Davies,
  Deller, Dettmar, van Diepen, Dijkstra, Donker, Doorduin, Dromer, Drost, van
  Duin, Eislöffel, van Enst, Ferrari, Frieswijk, Gankema, Garrett,
  de~Gasperin, Gerbers, de~Geus, Grießmeier, Grit, Gruppen, Hamaker, Hassall,
  Hoeft, Holties, Horneffer, van~der Horst, van Houwelingen, Huijgen,
  Iacobelli, Intema, Jackson, Jelic, de~Jong, Juette, Kant, Karastergiou,
  Koers, Kollen, Kondratiev, Kooistra, Koopman, Koster, Kuniyoshi, Kramer,
  Kuper, Lambropoulos, Law, van Leeuwen, Lemaitre, Loose, Maat, Macario,
  Markoff, Masters, McFadden, McKay-Bukowski, Meijering, Meulman, Mevius,
  Middelberg, Millenaar, Miller-Jones, Mohan, Mol, Morawietz, Morganti,
  Mulcahy, Mulder, Munk, Nieuwenhuis, van Nieuwpoort, Noordam, Norden, Noutsos,
  Offringa, Olofsson, Omar, Orrú, Overeem, Paas, Pandey-Pommier, Pandey,
  Pizzo, Polatidis, Rafferty, Rawlings, Reich, de~Reijer, Reitsma, Renting,
  Riemers, Rol, Romein, Roosjen, Ruiter, Scaife, van~der Schaaf, Scheers,
  Schellart, Schoenmakers, Schoonderbeek, Serylak, Shulevski, Sluman, Smirnov,
  Sobey, Spreeuw, Steinmetz, Sterks, Stiepel, Stuurwold, Tagger, Tang, Tasse,
  Thomas, Thoudam, Toribio, van~der Tol, Usov, van Veelen, van~der Veen, ter
  Veen, Verbiest, Vermeulen, Vermaas, Vocks, Vogt, de~Vos, van~der Wal, van
  Weeren, Weggemans, Weltevrede, White, Wijnholds, Wilhelmsson, Wucknitz,
  Yatawatta, Zarka, Zensus, \& van Zwieten}]{van_haarlem_lofar_2013}
van Haarlem, M.~P., Wise, M.~W., Gunst, A.~W., {et~al.} 2013, Astronomy and
  Astrophysics, 556, A2

\bibitem[{Wouthuysen(1952)}]{wouthuysen_excitation_1952}
Wouthuysen, S.~A. 1952, The Astronomical Journal, 57, 31

\end{thebibliography}

\end{document}